\documentclass[preprint,11pt]{elsarticle}
\usepackage{lineno,hyperref}
\usepackage{amssymb, amsmath,bm}
\usepackage{graphicx}
\usepackage{graphics}
\usepackage{latexsym}
\usepackage{booktabs}
\usepackage[utf8]{inputenc}
\usepackage{fullpage}
\journal{Physics Letters B}

\begin{document}

\begin{frontmatter}

\title{Microscopic calculation of the
%the first-forbidden low-Q-value 
$\beta^-$ decays of $^{151}$Sm, $^{171}$Tm, and $^{210}$Pb with implications to detection of the cosmic neutrino background}

\author{J.~Kostensalo}\address{Natural Resources Institute Finland, Yliopistokatu 6B, FI-80100 Joensuu, Finland}%

\author{J.~Kotila}
\address{Department of Physics, University of Jyv\"askyl\"a, P.O. Box 35, FI-40014, Jyv\"askyl\"a, Finland}
\address{Finnish Institute for Educational Research, University of Jyv\"askyl\"a, P.O. Box 35, FI-40014 Jyv\"askyl\"a, Finland}
\address{Center for Theoretical Physics, Sloane Physics Laboratory, Yale University, New Haven, Connecticut 06520-8120, USA}

\author{J.~Suhonen}
\address{Department of Physics, University of Jyv\"askyl\"a, P.O. Box 35, FI-40014, Jyv\"askyl\"a, Finland}

\begin{abstract}The electron spectral shapes corresponding to the low-$Q$ $\beta^-$-decay transitions
$^{151}\textrm{Sm}(5/2^-_{\rm g.s.})\to\,^{151}\textrm{Eu}(5/2^+_{\rm g.s.})$,
$^{151}\textrm{Sm}(5/2^-_{\rm g.s.})\to\,^{151}\textrm{Eu}(7/2^+_{1})$,
$^{171}\textrm{Tm}(1/2^+_{\rm g.s.})\to\,^{171}\textrm{Yb}(1/2^-_{\rm g.s.})$,
$^{171}\textrm{Tm}(1/2^+_{\rm g.s.})\to\,^{171}\textrm{Yb}(3/2^-_{1})$,
$^{210}\textrm{Pb}(0^+_{\rm g.s.})\to\,^{210}\textrm{Bi}(1^-_{\rm g.s.})$, and
$^{210}\textrm{Pb}(0^+_{\rm g.s.})\to\,^{210}\textrm{Bi}(0^-_{1})$ have been computed using beta-decay theory with several refinements for these first-forbidden nonunique (ff-nu) 
$\beta^-$ transitions. These ff-nu $\beta^-$ transitions have non-trivial electron spectral shapes with transition nuclear matrix elements (NMEs) computed by using the microscopic Interacting Boson-Fermion Model (IBFM-2) for the decays of $^{151}$Sm and $^{171}$Tm, and the nuclear shell model (NSM) for the decay of $^{210}$Pb. Within the respective $Q$ windows, the computed ff-nu electron spectral shapes deviate maximally at sub-percent level from the universal allowed shape, except for the transition
$^{210}\textrm{Pb}(0^+_{\rm g.s.})\to\,^{210}\textrm{Bi}(1^-_{\rm g.s.})$, where the maximal deviation is some 2.7$\%$. This confirms that the so-called $\xi$ approximation is fairly good for most of these low-$Q$ $\beta^-$ transitions and thus the allowed shape is a rather good first approximation. Our computed spectral shapes could be of interest for experiments aiming to measure the cosmic neutrino background (C$\nu$B), like the PTOLEMY experiment. We have also derived C$\nu$B cross sections for the ground-state transitions of the considered nuclei at the $\beta$ endpoint. Our findings indicate that more work on the atomic mismatch correction is needed in the future in order to extract reliable and precise C$\nu$B cross sections for any nuclear target.
\end{abstract}


\end{frontmatter}

%%  INTRO
Electron spectral shapes of forbidden nonunique $\beta$ decays can play a prominent role in several contexts of the present-day nuclear and particle physics, e.g., when trying to pin down the effective value $g_{\rm A}^{\rm eff}$ of the weak axial coupling in the context of $g_{\rm A}$-dependent spectral shapes (the Spectrum-Shape Method, SSM, introduced in \cite{Haaranen2016,Haaranen2017} and applied in \cite{Bodenstein2020,Kos2021}), and  when trying to explain the reactor antineutrino anomaly (RAA) \cite{Mention2011,Huber2011,Mueller2011} and the spectral "bump" related to the measured antineutrino flux from nuclear reactors \cite{Hayes2014,Fang2015,Hayen2019a,Hayen2019b,Berryman2020}. The problem of the effective value of $g_{\rm A}$ can have serious consequences for the sensitivity of the running and future experiments trying to detect the neutrinoless double beta decay \cite{Suhonen2017,Engel2017,Suhonen2019,Ejiri2019}. 

Another important context where the electron spectral shape of a forbidden nonunique 
$\beta$ transition plays a decisive role is the detection of the cosmic neutrino background (C$\nu$B) \cite{Gelmini2005,Cocco2007,Long2014}. The C$\nu$B is a relic of the early Universe and plays an essential role in understanding many key features of the microwave and dark-matter cosmology \cite{Dolgov2002,Les2006}. The wide evidence from cosmological surveys supports indirectly the existence of C$\nu$B, but direct evidence on the C$\nu$B is still lacking. Detection of C$\nu$B in controlled laboratory conditions would thus provide the first proof of the existence of non-relativistic neutrinos.

In the proposed experimental methods the detection of C$\nu$B leans on relic neutrino capture on an unstable but long-lived beta emitter with a sizable neutrino-capture cross section. In addition, a small decay energy ($Q$ value) is desirable in order to improve the detection potential of the experiment \cite{Brdar2022}. The PTOLEMY collaboration \cite{Betti2019} is actively considering to employ the decay of $^{171}$Tm \cite{Brdar2022} and in \cite{Cheipesh2021} the nuclei $^{151}$Sm and $^{171}$Tm are proposed as suitable candidates. Also the decay of the nucleus $^{210}$Pb is deemed interesting \cite{Brdar2022}. In addition, in \cite{Long2014}
a method to extract the needed relic-neutrino scattering cross section from the beta electron spectral shape has been presented. All the mentioned nuclei decay via first-forbidden nonunique (ff-nu) $\beta^-$ transitions to the ground state and the first excited state. These are the only final states in the $\beta$-decay $Q$ windows due to the smallness of the $Q$ values.

For low-$Q$-value ff-nu $\beta$ transitions in heavy nuclei the so-called $\xi$ approximation, where the associated electron spectral shape can be well approximated by the allowed shape, is usually valid \cite{Behrens1982, Mougeot2015}. In particular cases, dictated by the nuclear wave functions of the initial and final states, this approximation can be insufficient and a nuclear-structure calculation has to be done for the involved nuclear matrix elements (NMEs) \cite{Behrens1982, Mougeot2015,Kotani1959}. In \cite{Brdar2022} it was found indirectly that the $\xi$ approximation should be valid for the decays of $^{151}$Sm and $^{171}$Tm to their daughter nuclei $^{151}$Eu and $^{171}$Yb.
In this work we want to verify if this indirect method really gives reliable electron spectral shapes for these transition by calculating the NMEs involved in the transitions
$^{151}\textrm{Sm}(5/2^-_{\rm g.s.})\to\,^{151}\textrm{Eu}(5/2^+_{\rm g.s.})$,
$^{151}\textrm{Sm}(5/2^-_{\rm g.s.})\to\,^{151}\textrm{Eu}(7/2^+_{1})$,
$^{171}\textrm{Tm}(1/2^+_{\rm g.s.})\to\,^{171}\textrm{Yb}(1/2^-_{\rm g.s.})$, and
$^{171}\textrm{Tm}(1/2^+_{\rm g.s.})\to\,^{171}\textrm{Yb}(3/2^-_{1})$ by using a nuclear-structure model called the microscopic Interacting Boson-Fermion Model (IBFM-2) \cite{Iachello1991}, suitable for extracting wave functions of heavy, possibly deformed, nuclei. For the transitions
$^{210}\textrm{Pb}(0^+_{\rm g.s.})\to\,^{210}\textrm{Bi}(1^-_{\rm g.s.})$ and
$^{210}\textrm{Pb}(0^+_{\rm g.s.})\to\,^{210}\textrm{Bi}(0^-_{1})$ the involved NMEs have been calculated by using the nuclear shell model (MSM) \cite{Caurier2005}. This is possible owing to the (near) semi-magicity of the involved nuclei.

In addition to the $\beta$ spectral shapes we set out to determine the C$\nu$B scattering cross sections using the endpoint region of the computed $\beta$-electron spectra. The various corrective contributions to the cross sections are quantified and analyzed.

%% THEORY

% Beta spectrum and corrections

The beta-decay transitions discussed in this work are of the $\beta^-$ type and the 
corresponding half-life can be cast into the form
\begin{equation}
t_{1/2} = \frac{\kappa}{\tilde{C}},
\label{eq:hl}
\end{equation}
where $\kappa=6289\,\textrm{s}$ is a universal constant and $\tilde{C}$ is the so-called integrated shape function which is given by
\begin{equation}
\tilde{C} = \int_1^{w_0}F_0(Z,w_e)pw_e(w_0-w_e)^2K(Z,w_e)C(w_e)dw_e, 
\label{eq:cee}
\end{equation}
where $F_0(Z,w_e)$, is the Fermi function taking into account the final-state Coulomb distortion of the wave function of the emitted electron, $Z$ is the proton number of the final nucleus, and $w_0=W_0/m_e$, $w_e=W_e/m_e$, and $p=p_e/m_e=\sqrt{w_e^2-1}$ are dimensionless kinematic variables. Here $p_e$ and $W_e$ are the momentum and energy of the emitted electron, respectively, and $W_0$ is the beta endpoint energy. The factor $K(Z,w_e)$ includes all the correction terms. Here we follow the corrections discussed in \cite{Hayen2018} taking into account finite size, finite mass, radiative corrections, atomic exchange effects, and screening effects. The quantity of special interest in this work is the factor $C(w_e)$, known as the shape factor \cite{Behrens1982} and given by

%The shape factor is obtained by performing a multipole expansion of the $V-A$ hadronic current resulting in a complicated expression including momentum-dependent form factors, which are then expanded as power series (see \cite{Behrens1982}  for details). The shape-factor can be then written as  

\begin{align}
\notag
C(w_e)=&\sum_{k_e,k_{\nu},K}\lambda_{k_e}\Big\lbrack M_K(k_e,k_{\nu})^2 +m_K(k_e,k_{\nu})^2 \\
&-\frac{2\gamma_{k_e}\mu_{k_e}}{k_ew_e}M_K(k_e,k_{\nu})m_K(k_{e},k_{\nu})\Big\rbrack \,,
\label{eq:cwe}
\end{align}
where $k_{e}$ and $k_{\nu}$ come from the partial-wave expansion 
of the lepton wave functions, $\gamma_{k_e}=\sqrt{k_e^2-(\alpha Z)^2}$, $\mu_{k_e}\approx 1$, and 
$\lambda_{k_e}=F_{k_e-1}(Z,w_e)/F_0(Z,w_e)$ is the Coulomb function with $F_{k_e-1}(Z,w_e)$ being the 
generalized Fermi function. The quantities $M_K(k_e,k_{\nu})$ and $m_K(k_e,k_{\nu})$ have lengthy expressions which can be found from \cite{Behrens1982}. 

In the impulse approximation these form factors can be related to \emph{nuclear matrix elements} $\mathcal{M}_{KLs}^{(N)}$ by
\begin{align}
R^L \, ^VF^{(N)}_{KLs} &= (-1)^{K-L} g_V ^V\mathcal{M} ^{(N)}_{KLs} \\
R^L  \, ^AF^{(N)}_{KLs} &= (-1)^{K-L+1} g_A ^V\mathcal{M} ^{(N)}_{KLs}, 
\end{align}
where $R$ is the nuclear radius, and the sign convention is chosen as in, e.g., \cite{Haaranen2017}. The nuclear matrix elements can be expressed as
\begin{equation} 
^{V/A}\mathcal{M}_{KLs}^{(N)}=\frac{\sqrt{4\pi}}{\widehat{J}_i}
\sum_{pn} \, ^{V/A}m_{KLs}^{(N)}(pn)(\Psi_f|| [c_p^{\dagger}
\tilde{c}_n]_K || \Psi_i),
\label{eq:ME}
\end{equation}
where $^{V/A}m_{KLs}^{(N)}(pn)$ is the single-particle matrix element corresponding to the proton orbital $p$ and neutron orbital $n$, 
and $(\Psi_f|| [c_p^{\dagger}\tilde{c}_n]_K || \Psi_i)$ is the one-body 
transition density (OBTD), which contains the relevant nuclear-structure information. The choice of nuclear model enters the calculation through the evaluation of the OBTDs. 

% Xi approximation

The Behrens-B\"uhring formalism is based on expanding the matrix elements in the small quantities $W_eR$, $m_eR$, and $Z\alpha$ which allows for some power-series considerations. In the so-called $\xi$ approximation the complicated shape factor (\ref{eq:cwe}) is expressed in powers of $1/\xi$, where $\xi = Z\alpha/((W_e-m_e)R)$. It can then be shown that the shape factor of a forbidden transition is that of an allowed one with corrections of the order $Z\alpha/QR$, where $Q$ is the decay energy ($Q$ value) of the transition. When $1/\xi$ is small, allowed approximation of the spectrum shape is reasonably accurate. For the transitions considered here $\xi\approx 150$ for the ground-state-to-ground-state decay of $^{171}$Tm and even larger for the other transitions. Therefore, one would expect at most corrections of the order of 0.67\% to the spectra. However, as pointed out in e.g. \cite{Brdar2022}, there can be limitations to the applicability of the $\xi$ approach, in particular if notable cancellations appear among the various terms of the shape factor (\ref{eq:cwe}) \cite{Kotani1959}. In order to see whether such cancellations appear, a proper microscopic calculation of the shape factor must be performed with sufficient variation of the effective values of the involved weak-interaction parameters to get a realistic understanding of the involved uncertainties. Based on the studies \cite{Kubodera1978,Delorme1982,Warburton1991b,Kubodera1991,KostensaloPLB} we vary the value of the mesonic enhancement factor $\epsilon_{\rm MEC}$ of the axial-charge matrix element between 1.4 and 2.0, and based on the papers \cite{Haaranen2016,Haaranen2017,Kostensalo2017a,Kostensalo2017b} the value of the effective axial-vector coupling constant $g_{\rm A}^{\rm eff}$ between 0.80 and 1.20.

% capture rates

The neutrino capture rate at the beta endpoint can be derived from equations (\ref{eq:hl}) and (\ref{eq:cee}) and can be expressed as
\begin{equation}
    \bar{\sigma} = F_0(Z,w_e)pw_eK(Z,w_e)C(w_e) \big|_{w_e=w_0}.
    \label{eq:sigma}
\end{equation}
Relevant corrections here are the finite size, atomic screening, radiative, and atomic exchange corrections for which we adopt the expressions given in the comprehensive review \cite{Hayen2018} on allowed beta spectrum shape. The correction terms were evaluated at $W_e = Q - 0.01$ eV, up to which point they were stable in value. The full correction term is relatively stable up to this point. For forbidden transitions the shape factor is the most important correction. In addition to these corrections, there is one more non-trivial correction related to the shake-up and shake-off effects, where the final atom is either left in an excited state or ionized. For the heavy nuclei considered here, the shake-off probability is the dominant one, occurring for roughly 20--30\% of the decays \cite{Hayen2018,Carlson1970}. These effects can be reasonably accounted for with the so-called atomic mismatch correction, which is of the form \cite{Carlson1970,Desclaux1973} 
\begin{equation}
 r(Z,W_e) = 1 - \frac{1}{W_0-W_e} (44.2Z^{0.41}+2.3196\times10^{-7}Z^{4.45}) \, \rm eV + small \ correction.
 \label{eq:mis}
 \end{equation}

\begin{figure*}
%	\centering	
	\includegraphics[width=\textwidth]{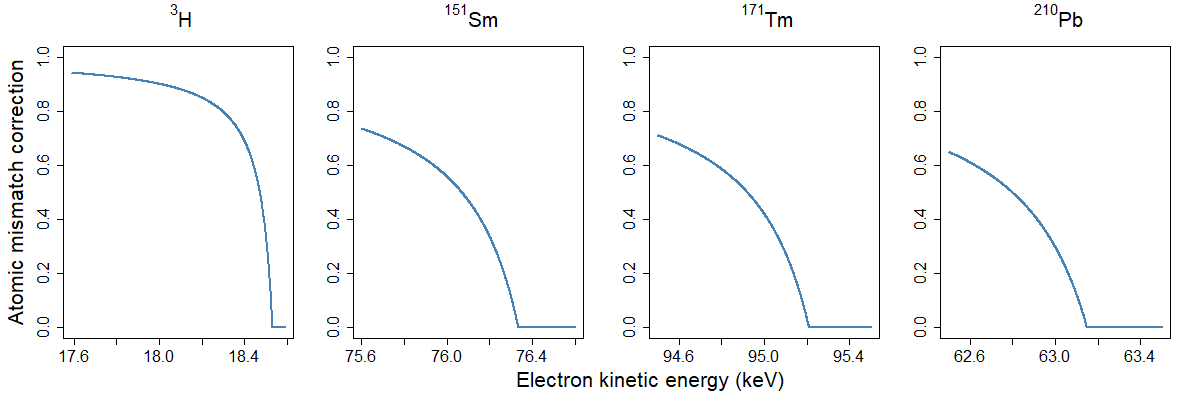}
	\caption{The mismatch correction of Eq.~(\ref{eq:mis}) for $^{3}$H and the other three C$\nu$B candidates studied in this work. 
\label{fig:mis}  }
	\end{figure*}

The correction (\ref{eq:mis}) is discussed in \cite{Hayen2018} in the context of spectral shapes, where its wild behavior near the endpoint is greatly mitigated by the kinematic term $(W_e - W_0)^2$ in Eq.~(\ref{eq:cee}) and is thus not problematic for the shape function and its integrated form (\ref{eq:cee}). However, when considering the cross section (\ref{eq:sigma}), evaluated at the beta endpoint, the correction is problematic, as it is quite small before rapidly going to zero very close to the endpoint. In \cite{Brdar2022} this difficulty was dealt with by fixing this correction to its value at $0.98Q$ for $W_e>0.98Q$, but this procedure is arbitrary and leaves room for improvements. In order to see the problem with this correction term, its behavior is presented for $^{3}$H, a considered candidate for the C$\nu$B detection \cite{Long2014}, and the three candidates studied in this work in Fig.~\ref{fig:mis} for the energy range $W_e = [Q - 1\,{\rm keV}\,,Q]$. The $Q$ value is now shifted downwards by the mean atomic excitation energy. Also, the cross section near the new lower endpoint goes rapidly to zero. Since in reality the shake-up and shake-off processes have some finite probability (smaller than one) of occurring, the $Q$-value shift would only happen for some fraction of the transitions. Thus, it is clear that for the C$\nu$B detection this effect needs to be considered in a more sophisticated way. Since details of the atomic corrections are out of the scope of the present paper we leave this correction out from the present calculations and concentrate on nuclear structure of the shape factor and the other well-behaved corrections. 

%  IBFM-2 theory
The shape factors related to the decay transitions in $^{151}$Sm and $^{171}$Tm were treated in the nuclear-structure framework of the microscopic Interacting Boson-Fermion Model (IBFM-2) \cite{Iachello1991}. IBFM-2 is an extension of the well-known microscopic Interacting Boson Model (IBM-2) \cite{Arima:1976ky,Iachello:2006fqa} to odd-mass nuclear systems.
The IBM-2 is a phenomenological approach that has been one of the most successful models in reproducing collective features of the low-lying levels of medium-heavy as well as heavy nuclei. The IBM-2 deals with even-even nuclei, where one replaces valence-nucleon pairs with bosons with angular momentum 0 or 2. By coupling an extra fermion to this bosonic system, one is able to extend the IBM-2 to the study of odd-$A$ nuclei. This extension is the IBFM-2. 

The mapping of the single-fermion creation operator onto the IBFM-2 space follows the procedure introduced in Ref. \cite{barea14,Matus:2017eni} where relevant terms, using exact values for the fermion matrix elements in the Generalized Seniority scheme, were worked out and thus use of the Number Operator Approximation (NOA) was avoided. This method has already been applied to allowed beta decays in Ref. \cite{Mardones:2016wgy} and now we extend its use to ff-nu beta decays in the $^{151}$Sm and $^{171}$Tm nuclei. Since these nuclei are mid-shell nuclei they are best described by IBFM-2. Contrary to this, the nuclei $^{210}$Pb and $^{210}$Bi, for which the IBFM-2 model is not applicable since they are even-$A$ nuclei, are in the vicinity of the doubly-magic nucleus $^{208}$Pb, and thus can be best described using the nuclear shell model. 

% IBFM-2 application
In the  IBFM-2 calculations the even-even $^{150}$Sm nucleus was used as a common core for the odd $^{151}$Sm and $^{151}$Eu nuclei, and $^{170}$Yb and $^{172}$Yb were adopted as cores for the $^{171}$Yb and $^{171}$Tm nuclei, respectively. The parameters for the core Sm and Yb nuclei were taken from Refs.~\cite{scho1980,Hady_2020}, respectively. The valence space was chosen to span $2p, 1f, 0h_{9/2}, 0i_{13/2}$  neutron and $2s, 1d, 0g, 0h_{11/2}$ proton orbitals with unperturbed single-particle energies taken from \cite{kotila16}, where the effect of single-particle energies on occupation probabilities was studied. The used boson-fermion interaction parameters are listed in Table~\ref{table:ibfm_par}.

\begin{table}
\caption{\label{table:ibfm_par} Boson-fermion interaction parameters (in MeV) used in the IBFM-2 calculations.}
\begin{center}
\begin{tabular}{cccc}
\hline
			&$\Gamma_\rho$ &$\Lambda_\rho$ &$A_\rho$\\
\hline
$^{151}$Sm	&-0.400	&0.050	&-0.700	\\
$^{151}$Eu	&-0.080	&0.009	&-0.050	\\
$^{171}$Tm	&-0.050	&-0.027	&0.580	\\
$^{171}$Yb	&-0.062	&-0.021	&0.200	\\
\hline
\end{tabular}
\end{center}
\end{table}

% Shell model application

The wave functions and one-body densities for the decay of $^{210}$Pb were calculated in the shell-model framework using the computer program NuShellX@MSU \cite{nushellx}. The calculations were done in the full model space spanning the proton orbitals $0h_{9/2},2p,1f,0i_{13/2}$ and the neutron orbitals $0i_{11/2},1g,2d,3s,0j_{15/2}$ using the effective Hamiltonian \emph{khpe} \cite{Warburton1991}.

%% RESULTS AND DISCUSSION

			\begin{figure*}
%	\centering	
	\includegraphics[width=\textwidth]{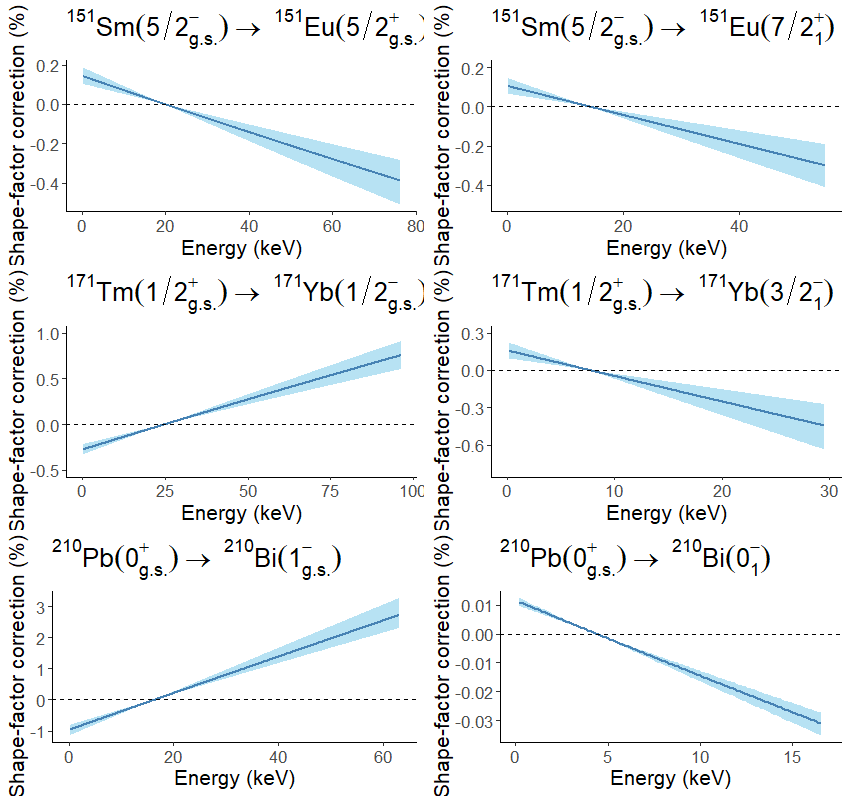}
	\caption{Shape-factor corrections to the allowed shape of the normalized electron spectrum and the related uncertainties for each transition discussed in this work. The uncertainties have been obtained by varying the effective axial-vector coupling constant $g_{\rm A}^{\rm eff}$ between 0.80 and 1.20 and the mesonic enhancement $\epsilon_{\rm MEC}$ of the axial-charge matrix element between 1.4 and 2.0.
\label{fig:corr}  }
	\end{figure*}

The calculated shape factors for the decays of $^{151}$Sm, $^{171}$Tm, and $^{210}$Pb are compared with the allowed approximation in Fig. \ref{fig:corr}. For all six transitions the corrections to the allowed spectral shape are largest at the endpoint, which is the region of interest for detecting the cosmic neutrino background. The endpoint corrections are given in Table \ref{table:endpoint}. While the pure pseudoscalar transition $^{210}\rm Pb(0^+_{\rm g.s.}) \rightarrow \, ^{210}\rm Bi(0^-_1)$ gets only a tiny correction of $-0.031\pm0.004$\% the transition $^{210}\rm Pb(0^+_{\rm g.s.}) \rightarrow \, ^{210}\rm Bi(1^-_{\rm g.s.})$ gets a non-trivial $2.72_{-0.42}^{+0.54}$\% correction. The shape-factor corrections for the four other transitions are roughly 0.50\%. Looking at the dominating ground-state-to-ground-state transitions the corrections are 0.71$/\xi$ for $^{151}$Sm, 1.16$/\xi$ for $^{171}$Tm, and 7.10$/\xi$ for $^{210}$Pb. In addition, for the three decays to the excited states the corrections are roughly $1/\xi.$ While these results are consistent with the $\mathcal{O}(1/\xi)$, the results highlight the fact that one should not assume that the error is necessarily exactly $\approx 1/\xi$. The decay of $^{210}$Pb to the ground state is a good example of destructive interference, where the $\xi$-approximation does not hold very well. 

\begin{table}
\begin{center}
\caption{\label{table:endpoint} Beta-spectrum endpoint corrections and their uncertainties. The uncertainties have been obtained by varying the effective axial-vector coupling constant $g_{\rm A}^{\rm eff}$ between 0.80 and 1.20 and the mesonic enhancement $\epsilon_{\rm MEC}$ of the axial-charge matrix element between 1.4 and 2.0. }
\begin{tabular}{ccc}
\hline
	Nucleus		&$J^\pi_f$ &Endpoint correction (\%)\\
\hline
$^{151}$Sm	&	5/2$^+$	&		$-0.39_{-0.12}^{+0.10}$	\\
	&	7/2$^+$	&		$-0.30\pm0.10$	\\
$^{171}$Tm 	&	1/2$^-$	&		$0.75\pm0.15$	\\
	&	3/2$^-$	&		$-0.44_{-0.19}^{+0.17}$	\\
$^{210}$Pb	&	0$^-$	&	$-0.031\pm0.004$	\\
	&	1$^-$	&		$2.72_{-0.42}^{+0.54}$	\\

\hline
\end{tabular}
\end{center}
\end{table}

\begin{table}
\begin{center}
\caption{\label{table:sigma} Endpoint cross sections and their leading errors for the ground-state transitions of the discussed mother nuclei. All relevant corrections except the atomic mismatch effect are included. }
\begin{tabular}{cccccc}
\hline
	Nucleus			&	$\bar{\sigma}$ (cm$^2$)	&	Half-life error	&	Q-value error	&	Spectrum-shape error	&	Total error 		\\
	\hline
$^{151}$Sm	&	4.79$\times10^{-48}$	&	0.44$\times10^{-48}$	&	0.01$\times10^{-48}$	&	0.01$\times10^{-48}$ & 0.44$\times10^{-48}$	\\
$^{171}$Tm	&	1.14$\times10^{-46}$ 	&	0.01$\times10^{-46}$	&	0.07$\times10^{-46}$	&	0.01$\times10^{-46}$	&	0.07$\times10^{-46}$	\\
$^{210}$Pb	&	3.27$\times10^{-48}$	&	0.04$\times10^{-48}$	&	0.02$\times10^{-48}$	&	0.01$\times10^{-48}$	&	0.05$\times10^{-48}$	\\
\hline
\end{tabular}
\end{center}
\end{table}

The cross sections and the impacts of the relevant corrections for the ground-state-to-ground-state transitions of $^{151}$Sm, $^{171}$Tm, and $^{210}$Pb are given in Table \ref{table:sigma}. For $^{151}$Sm and $^{171}$Tm the obtained values $(4.79\pm0.44)\times10^{-48}\, \rm cm^2$ and $(1.14\pm0.07)\times10^{-46}\, \rm cm^2$ are somewhat larger than the values $(4.77\pm0.01)\times10^{-48}\, \rm cm^2$ and $(1.12\pm0.01)\times10^{-46}\, \rm cm^2$ reported in \cite{Brdar2022}. The small deviations in the estimates are due to some differences in how the correction terms are evaluated \cite{BrdarPrivate} and the larger uncertainties in the present work are due to the inclusion of all the relevant sources of uncertainty.

%% CONCLUSIONS
In this Letter we have calculated the endpoint cross sections for $^{151}$Sm, $^{171}$Tm, and $^{210}$Pb using realistic microscopic nuclear models for the involved wave functions. These low-$Q$-value decays are potential candidates for the detection of cosmic neutrino background. Out of these candidates the most promising one is $^{171}$Tm with $\bar{\sigma} =(1.14\pm0.07)\times10^{-46}\, \rm cm^2$. The validity of the $\xi$ approximation for forbidden spectral shapes was investigated and errors up to 7.10$/\xi$ were recorded. While this can be considered consistent with an $\mathcal{O}(1/\xi)$ estimate, our results highlight the fact that one should not assume a-priori an error $\approx 1/\xi$, but nuclear structure can alter the situation depending on the details of the initial and final nuclear wave functions.

\textbf{Declaration of competing interest}

The authors declare that there are no known competing financial interests or personal relationships that could have appeared to influence the work reported in this paper.

\textbf{Acknowledgements}

This work was supported by the Academy of Finland, Grant Nos. 314733, 345869 and 318043.


\begin{thebibliography}{00} 
\bibitem{Haaranen2016} M. Haaranen, P. C. Srivastava, J. Suhonen, Phys. Rev. C 93 (2016) 034308.
\bibitem{Haaranen2017} M. Haaranen, J. Kotila, J. Suhonen, Phys. Rev. C 95 (2017) 024327.
\bibitem{Bodenstein2020} L. Bodenstein-Dresler \textit{et al.} (COBRA Collaboration), Phys. Lett. B 800 (2020) 135092.
\bibitem{Kos2021} J. Kostensalo, J. Suhonen, J. Volkmer, S. Zatschler, K. Zuber, Phys. Lett. B 822 (2021) 136652.
\bibitem{Mention2011} G. Mention, M. Fechner, T. Lasserre, T. A. Mueller, D. Lhuillier, M. Cribier, A. Letourneau, Phys. Rev. D 83 (2011) 073006.
\bibitem{Huber2011} P. Huber, Phys. Rev. C 84 (2011) 024617.
\bibitem{Mueller2011} T. A. Mueller \textit{et al.}, Phys. Rev. C 83 (2011) 054615.
\bibitem{Hayes2014} A. C. Hayes, J. L. Friar, G. T. Garvey, G. Jungman, G. Jonkmans, Phys. Rev. Lett. 112 (2014) 202501.
\bibitem{Fang2015} D. L. Fang, B. A. Brown, Phys. Rev. C 91 (2015) 025503.
\bibitem{Hayen2019a} L. Hayen, J. Kostensalo, N. Severijns, J. Suhonen, Phys. Rev. C 99 (2019) 031301(R).
\bibitem{Hayen2019b} L. Hayen, J. Kostensalo, N. Severijns, J. Suhonen, Phys. Rev. C 100 (2019) 054323.
\bibitem{Berryman2020} J. M. Berryman, P. Huber, Phys. Rev. D 101 (2020) 015008.
\bibitem{Suhonen2017} J. T. Suhonen, Front. Phys. 5 (2017) 55.
\bibitem{Engel2017} J. Engel, J. Men{\' e}ndez, Rep. Prog. Phys. 60 (2017) 046301.
\bibitem{Suhonen2019} J. Suhonen, J. Kostensalo, Front. Phys. 7 (2019) 29.
\bibitem{Ejiri2019} H. Ejiri, J. Suhonen, K. Zuber, Phys. Rep. 797 (2019) 1.
\bibitem{Gelmini2005} G. B. Gelmini, Physica Scripta T 121 (2005) 131. 
\bibitem{Cocco2007} A. G. Cocco, G. Mangano, M. Messina, J. Cosmol. Astropart. Phys. 06 (2007) 015. 
\bibitem{Long2014} A. J. Long, C. Lunardini, E. Sabancilar, J. Cosmol. Astropart. Phys. 08 (2014) 038. 
\bibitem{Dolgov2002} A. Dolgov, Phys. Rep. 370 (2002) 333.
\bibitem{Les2006} J. Lesgiurgues, S. Pastor, Phys. Rep. 429 (2006) 307.
\bibitem{Brdar2022} V. Brdar, R. Plestid, N. Rocco, Phys. Rev. C 105 (2022) 045501.
\bibitem{Betti2019} M. G. Betti, et al., PTOLEMY Collaboration, J. Cosmol. Astropart. Phys. 07 (2019) 047. 
\bibitem{Cheipesh2021} Y. Cheipesh, V. Cheianov, A. Boyarsky, Phys. Rev. D 104 (2021) 116004.
\bibitem{Behrens1982} H. Behrens, W. B\"uhring, \textit{Electron Radial Wave Functions and Nuclear Beta Decay} (Clarendon, Oxford, 1982).
\bibitem{Mougeot2015} X. Mougeot, Phys. Rev. C 91 (2015) 055504.
\bibitem{Kotani1959} T. Kotani, Phys. Rev. 114 (1959) 795.
\bibitem{Iachello1991} F. Iachello, P. Van Isacker, {\it The Interacting Boson-Fermion Model} (Cambridge University Press, 1991). 
\bibitem{Caurier2005} E. Caurier, G. Mart{\'i}nez-Pin{\'e}do, F. Nowacki, A. Poves, A. P. Zuker, Rev. Mod. Phys. 77 (2005) 427. 
\bibitem{Hayen2018} L. Hayen, N. Severijns, K. Bodek, D. Rozpedzik, X. Mougeot, Rev. Mod. Phys. 90 (2018) 015008.
\bibitem{Kubodera1978} K. Kubodera, J. Delorme, M. Rho, Phys. Rev. Lett. 40 (1978) 755.
\bibitem{Delorme1982} J. Delorme, Nucl. Phys. A374 (1982) 541c.
\bibitem{Warburton1991b} E. K. Warburton, Phys. Rev. C 44 (1991) 233.
\bibitem{Kubodera1991} K. Kubodera, M. Rho, Phys. Rev. Lett. 67 (1991) 3479.
\bibitem{KostensaloPLB} J. Kostensalo, J. Suhonen, Phys. Lett. B 781 (2018) 480.
\bibitem{Kostensalo2017a} J. Kostensalo, M. Haaranen, and J. Suhonen, Phys. Rev. C 95 (2017) 044313.
\bibitem{Kostensalo2017b} J. Kostensalo and J. Suhonen, Phys. Rev. C 96 (2017) 024317.
\bibitem{Carlson1970} T.A. Carlson, C. Nestor, N. Wasserman, J. Mcdowell, At. Data Nucl. Data Tables 2 (1970) 63.
\bibitem{Desclaux1973} J.P. Desclaux, At. Data Nucl. Data Tables 12(4) (1973) 311.
\bibitem{Arima:1976ky} A. Arima, F. Iachello, Ann. Phys. 99 (1976) 253; 111 (1978) 201; 123 (1979) 468.  
\bibitem{Iachello:2006fqa} F. Iachello, A. Arima, {\it The Interacting Boson Model} (Cambridge University Press, 1987).
\bibitem{barea14} J. Barea, C. E. Alonso, J. M. Arias, Phys. Lett. B 737 (2014) 205.
\bibitem{Matus:2017eni} F. A. Matus, J. Barea, Phys. Rev. C 95 (2017) 034317.
\bibitem{Mardones:2016wgy} E. Mardones, J. Barea, C. E. Alonso, J. M. Arias, Phys. Rev. C 93 (2016) 034332.
\bibitem{scho1980} O. Scholten, Ph.D. thesis (University of Groningen, The Netherlands, 1980).
\bibitem{Hady_2020} H. N Hady and M. K. Muttal, J. Phys.: Conf. Ser. 1591 (2020) 012016.
\bibitem{kotila16} J. Kotila, J. Barea, Phys. Rev. C. 94 (2016) 034320.
\bibitem{nushellx} B. A. Brown, W. D. M. Rae, Nucl. Data Sheets 120 (2014) 115.
\bibitem{Warburton1991} E. K. Warburton, B.A. Brown, Phys. Rev. C 43 (1991) 602.
\bibitem{Mikulenko2021} O. Mikulenko, Y. Cheipesh, V. Cheianov, A. Boyarsky, DOI:10.48550/arXiv.2111.09292.
\bibitem{BrdarPrivate} V. Brdar (private communication).
\bibitem{NNDC} National Nuclear Data Center (NNDC), NuDat 3.0 https://www.nndc.bnl.gov/nudat3/ (cited 6/19/2022).
\end{thebibliography}
\end{document}